# Resonant Self-Diffraction of Femtosecond Extreme Ultraviolet Pulses in Cobalt

Alexei A. Maznev,[1,*] Wonseok Lee,[2] Scott K. Cushing,[2] Dario De Angelis,[3] Danny Fainozzi,[3] Laura Foglia,[3] Christian Gutt,[4] Nicolas Jaouen,[5,6] Fabian Kammerbauer,[7] Claudio Masciovecchio,[3] Riccardo Mincigrucci,[3] Keith A. Nelson,[1] Ettore Paltanin,[3] Jacopo Stefano Pelli-Cresi,[3] Vincent Polewczyk,[8] Dmitriy Ksenzov,[4] Filippo Bencivenga.[3]

[1]Department of Chemistry, Massachusetts Institute of Technology, Cambridge, Massachusetts 02139, USA
[2]Division of Chemistry and Chemical Engineering, California Institute of Technology, California 91125, USA
[3]Elettra-Sincrotrone Trieste, SS 14 km 163,5 in AREA Science Park, 34149 Trieste, Italy
[4]Department Physik, Universität Siegen, Walter-Flex-Strasse 3, 57072 Siegen, Germany
[5]Synchrotron SOLEIL, L'Orme des Merisiers, Saint-Aubin, Gif-sur-Yvette Cedex, 91192, France
[6]Department of Molecular Sciences and Nanosystems, Ca' Foscari University of Venice, 30172 Venezia, Italy
[7]Institute of Physics, Johannes Gutenberg University Mainz, 55099 Mainz, Germany
[8]Université Paris-Saclay, UVSQ, CNRS, GEMaC, 78000, Versailles, France
*Corresponding author: alexei.maznev@gmail.com

**Abstract**
Self-diffraction is a non-collinear four-wave mixing technique well-known in optics. We explore self-diffraction in the extreme ultraviolet (EUV) range, taking advantage of intense femtosecond EUV pulses produced by a free electron laser. Two pulses are crossed in a thin cobalt film and their interference results in a spatially periodic electronic excitation. The diffraction of one of the same pulses by the associated refractive index modulation is measured as a function of the EUV wavelength. A sharp peak in the self-diffraction efficiency is observed at the $M_{2,3}$ absorption edge of cobalt at 59 eV and a fine structure is found above the edge. The results are compared with a theoretical model assuming that the excitation results in an increase of the electronic temperature. EUV self-diffraction offers a potentially useful spectroscopy tool and will be instrumental in studying coherent effects in the EUV range.

The advent of free electron lasers (FELs) enabled the expansion of nonlinear optical spectroscopy methods into extreme ultraviolet (EUV) and X-ray ranges [1]. In particular, four-wave mixing (FWM) techniques with EUV and EUV/optical fields are being actively developed [2–5]. Self-diffraction (SD) is the simplest non-collinear FWM process, in which the interference of two coherent beams crossed in the sample results in a spatially periodic excitation acting as a diffraction grating for the same beams. It is a well-known technique [6,7] widely used in nonlinear optical studies [8–11]. The SD geometry has also been used by theoreticians investigating nonlinear optical interactions in condensed matter [12, 13].

In the EUV range, a related FWM technique referred to as EUV transient grating (TG) spectroscopy has recently been developed at the FERMI FEL [4,14]. In the TG technique, a time-delayed probe pulse diffracts off a spatially periodic excitation produced by two time-coincident pump pulses crossed at the sample. SD of time-coincident pulses can be thought of as a degenerate version of a TG experiment at a zero pump-probe delay, where a pump beam serves as a zero-delay probe. The SD geometry has the advantage of simplicity, which is important at short wavelengths where manipulating multiple noncollinear beams is significantly more difficult [15] than in a conventional optical experiment. Even more importantly, in the EUV TG experiments performed to date, the FEL wavelength could not be continuously tuned because of the narrow-band multilayer mirrors in the probe beam path. The SD approach overcomes this limitation, enabling studies in the vicinity of resonances, which has been identified by theoreticians as an especially promising application of FWM at short wavelengths [16].

In this report, we describe an EUV SD experiment on a thin cobalt film, with the photon energy scanned across the $M_{2,3}$ edge of Co. We observe a great enhancement of the SD efficiency at the resonant absorption edge and a fine structure above the edge. The results are compared with *ab initio* calculations using density functional theory (DFT) and the Bethe-Salpeter equation (BSE) based on the assumption that the electronic system is nearly thermalized within the FEL pulse duration.

The experiment was conducted at the TIMER beamline at FERMI and is schematically shown in Fig. 1(a). Two 50 fs EUV pulses are obtained by bisecting the FEL output with a grazing incidence mirror. The beams are spatially and temporally overlapped at the sample, with a crossing angle of $2\theta = 18.4°$ (the bisector being normal to the sample surface) and a FWHM spot size of 300 μm. Each of the incident beams gives rise to two first order SD beams. One of the SD beams from pump A coincides with the transmitted pump B; the other SD beam goes into a background-free direction making an angle of $\psi = \arcsin(3\sin(\theta)) = 28.7°$ to the sample normal and is detected by a CCD camera [17]. The sample was a 20-nm-thick Co film deposited by e-beam evaporation on a 30 nm silicon nitride membrane. The measurements were performed at room temperature in high vacuum.

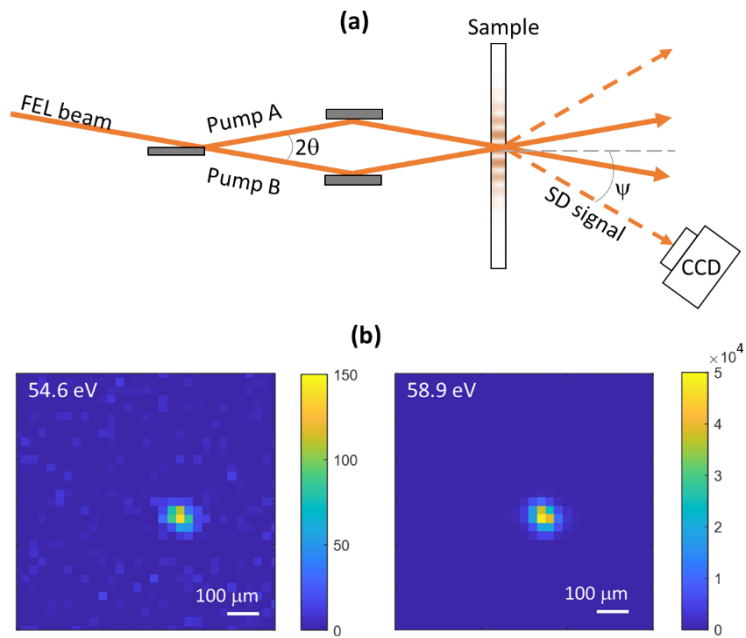

FIG. 1. (a) Schematic of the experiment. Dashed lines show background-free SD beams. (b) Images of the SD signal spot on the detector at photon energies corresponding to the smallest (54.6 eV) and largest (58.9 eV) SD signal. The color scale bars show CCD counts (note the substantial scale difference between the two images).

The CCD camera images obtained by averaging over 250 FEL shots show a bright spot at the expected position of the SD signal, which only appears when the two pump beams are overlapped at the sample, see representative images in Fig. 1(b). To quantify the SD signal, the CCD image is integrated within a region of interest chosen to include the entire SD spot. Figure 2(a) shows the dependence of the SD signal on the FEL photon energy ($h\nu$) in the range 54.6 – 72.1 eV (wavelengths 17.2 – 22.7 nm). For each value of $h\nu$, we obtained 3 – 8 images used to calculate the average value shown in Fig. 2(a) and the standard error shown as the error bars. The FEL pulse energy at the sample was ~1 μJ, with variations within a factor of two across the $h\nu$ scan. To compensate for the variations of the FEL intensity, the signal is normalized by the cube of the FEL intensity in accordance with the expected intensity dependence of a FWM signal.

The most conspicuous feature in Fig. 2(a) is the sharp increase of the SD signal at the absorption edge. Between 54.6 eV and the peak at 58.8 eV, the increase is a factor of 500. Past the absorption edge, the SD signal decreases but remains higher than pre-edge. Furthermore, the SD spectrum above the edge exhibits a fine structure involving a shoulder at ~61 eV, a dip at 62.5 eV, and a peak at 64 eV. This fine structure is not present in the EUV absorption spectrum of Co [18].

Figure. 2(b) shows the dependence of the SD signal on the FEL pulse energy at the sample at 59 eV, i.e., near the peak of the SD response, and confirms the cubic dependence mentioned above [19]. Note that the dynamic range of this measurement amounted to 4 orders of magnitude and the signal level, on the high end, exceeded 5000 photons/shot.

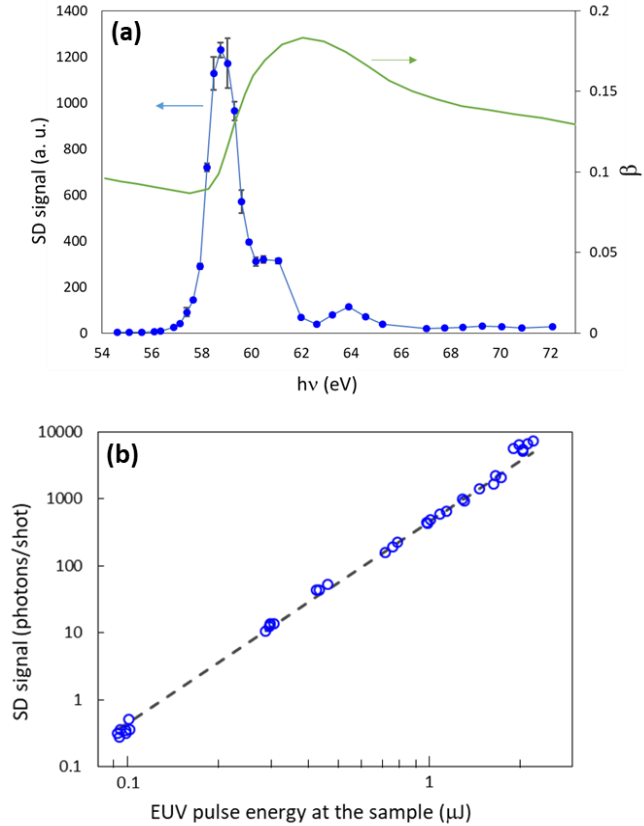

FIG. 2. (a) The dependence of the SD signal normalized by the cube of the FEL intensity on the photon energy (blue) and the spectrum of the imaginary part of the refractive index $\beta$ adopted from Ref. [18] (green). Error bars reflect the standard error obtained from multiple measurements done at the same photon energy; where the error bars are not shown, they are smaller than the symbol size. (b) The dependence of the SD signal on the FEL intensity at the sample at $h\nu = 59$ eV (symbols). The dashed line shows a cubic dependence fit.

A natural question to ask is whether the observed resonant SD phenomenon is due to a coherent $\chi^{(3)}$ response involving the core hole and initially excited electron [16], or whether it is caused by an incoherent population of excited electrons resulting from multi-step relaxation of the initial excitation [20]. The dephasing time of the $M_2$ and $M_3$ resonances in Co is estimated from the corresponding linewidths [21] as ~3 fs, which is much shorter than the FEL pulse duration. While the short dephasing time does not preclude the presence of a coherent $\chi^{(3)}$ response, the latter seems inconsistent with a recently reported TG measurement with a variably delayed probe pulse on a similar sample [22] with the pump and probe photon energy 59.6 eV, i.e., close to the peak of our SD spectrum. In that experiment, the initial fast response was characterized by a short rise time comparable to the FEL pulse duration and a slower decay time of ~250 fs. Due to the short dephasing time, a coherent $\chi^{(3)}$ response would only exist while the pump and probe pulses overlapped, which is incompatible with the observed slow decay. As mentioned above, our SD measurement can be considered a degenerate version of TG experiment at the zero pump-probe delay. Consequently, we believe that our SD signal is more likely to be produced by an incoherent population of excited electrons than by a coherent $\chi^{(3)}$ response.

The simplest way to model the formation of the SD signal due to an incoherent electronic excitation is to assume that the electrons get thermalized faster than the FEL pulse duration. It was reported that electronic thermalization following optical excitation of Co occurs within 2 fs [23]. While thermalization of the excitation produced by ~60 eV EUV photons may take longer, we believe that assuming a thermalized electronic sub-system is a reasonable first step towards developing the theory of the observed phenomenon.

We calculate the variation of the complex refractive index at the EUV wavelength caused by an electronic temperature change using DFT and BSE [24–27]. The electronic temperature is implemented in the BSE Hamiltonian $H^{BSE}$ by modifying the fractional occupation numbers [28–30]

$$H_{ij}^{BSE} = \epsilon_i \delta_{ij} + \sqrt{\widetilde{f_i}}[V_X - W]\sqrt{\widetilde{f_j}}. \tag{1}$$

In Eq. (1), the notation $i = \{vc\mathbf{k}\}$ represents the electron-hole pair indexes for each valence and conduction bands ($v$ and $c$) and k-points ($\mathbf{k}$). The Hamiltonian includes the bare energies $\epsilon_i$, the occupation number difference between the nominal electron ($f_{ei}$) and hole ($f_{hi}$) states $\widetilde{f_i} = |f_{ei} - f_{hi}|$, and the two electron-hole interaction terms, the direct $W$ and exchange $V_X$. While the electron-hole correlation function in the BSE can be expanded into an infinite series, the first-order electron-hole interaction terms are only considered here. Additionally, the Tamm-Dancoff approximation is employed to simplify the calculations and make them computationally more efficient, while still providing reasonably accurate results for calculating the complex dielectric function [31]. The individual occupation numbers are given by the Fermi-Dirac distribution function $(E, \mu(T_e), T_e) = \{\exp[(E - \mu(T_e))/(k_B T_e)] + 1\}^{-1}$ where $\mu$ is the chemical potential, $k_B$ is the Boltzmann constant, and $T_e$ indicates the electron temperature. The chemical potential is determined using the electronic density of states in Co obtained from DFT [32]. By solving the BSE via the Haydock recursion method, the complex dielectric function is derived from the photon operator $\hat{T}$ acting on the ground state $|\Phi_0\rangle$ and two-particle Green's function at energy $\omega$ defined as $G_2(\omega) = (\omega - H^{BSE} + i\eta)^{-1}$ with the BSE Hamiltonian and broadening parameter $\eta$,

$$\epsilon(\omega) = 1 - \frac{4\pi}{\Omega_V q^2}\left\langle\Phi_0\left|\hat{T}^\dagger \frac{1}{\omega - H^{BSE} + i\eta}\hat{T}\right|\Phi_0\right\rangle, \tag{2}$$

where $\Omega_V$ is the unit cell volume and $q$ denotes the magnitude of the photon momentum [26, 33]. The dielectric function is then converted into the complex refractive index: $\tilde{n} = 1 - \delta + i\beta$.

The diffraction efficiency of a refractive index grating in a weakly absorbing material is proportional to $(\Delta\delta)^2 + (\Delta\beta)^2$, where $\Delta\delta$ and $\Delta\beta$ are the amplitudes of the modulation of the real and imaginary parts of the refractive index, respectively [34]. In our case, the material is strongly absorbing at the EUV wavelength, which leads to a more complicated expression for the diffraction efficiency [22]. Representing the refractive index variation as $(\Delta\delta)^2 + (\Delta\beta)^2 = A(h\nu)\rho^2$, where $A(h\nu)$ is a function of the photon energy and $\rho$ is the absorbed energy density and using Eq. (5) from Ref. [22], we obtain the following expression for the normalized SD signal,

$$\frac{I_{SD}}{I_0^3} \propto A(h\nu)\frac{e^{-\frac{d}{L\cos\psi}}}{L^2} \frac{e^{-\frac{d}{L^*}} - 2e^{-\frac{d}{2L^*}}\cos(\Delta Q_z d) + 1}{\Delta Q_z^2 + \frac{1}{4L^{*2}}}, \tag{3}$$

where $I_{SD}$ is the SD signal, $I_0$ is the pump pulse energy, $d$ is the sample thickness, $L$ is the absorption length at the EUV wavelength, $L^* = L(3/\cos\theta - 1/\cos\psi)^{-1}$, and $\Delta Q_z = k(\cos\theta - \cos\psi)$, where $k$ is the EUV wave vector. (We only retained the terms dependent on the photon energy.) We calculate $A(h\nu)$ by computing $(\Delta\delta)^2 + (\Delta\beta)^2$ at a fixed electronic temperature rise of 100 K above a background temperature of 300 K [32] and take $L(h\nu)$ from Ref. [18]. (As one can see from the supplemental Fig. S4 [20], the variations of $\delta$ and $\beta$ are comparable in magnitude.)

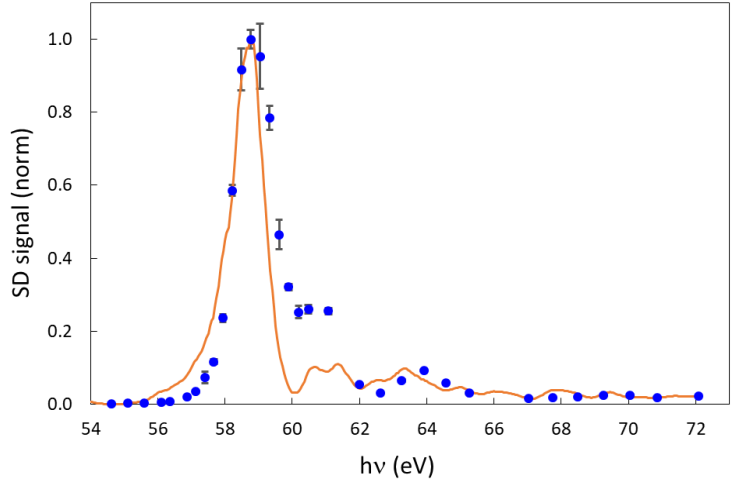

FIG. 3. Calculated SD spectrum (solid curve) based on the assumption that the EUV excitation modulates the complex refractive index via electronic temperature vs experimental data (dots). Both the theoretical curve and experimental data are normalized to unity at their maxima.

Figure 3 shows the calculated SD spectrum alongside the experimental data. The calculated SD spectrum reproduces a sharp peak at the Co $M_{2,3}$ edge and exhibits smaller peaks above the edge, qualitatively agreeing with experiment. At the Co $M_{2,3}$ edge, EUV photons excite core 3p electrons to unoccupied states near the Fermi level in the conduction band formed by 3d states. Changes in the electron temperature alter the electronic population near the Fermi level, as described by Fermi-Dirac statistics. When the final state of the transition lies at the Fermi level, the refractive index becomes highly sensitive to variations in the electronic temperature, which explains the intense SD peak. The interactions between electrons involved in the transition and screening processes lead to collective excitations above the Co $M_{2,3}$ edge [35]. This many-body effect is captured by using the screened Coulomb potential in the direct term ($W$) of the BSE Hamiltonian in Eq. (1). The transition probability is modified due to the changes in the electronic population and spectral changes resulting from the state-filling effect are therefore observed beyond the absorption edge. The fine structure above the edge is less accurately reproduced by our calculations, because the model used here is primarily designed for calculating the spectrum near an absorption edge and may not be as suitable for simulating the extended SD spectrum.

While the present experiment provides a proof-of-principle for resonant EUV SD, further developments can be anticipated. Firstly, the fact that the M-edge resonance is much more prominent in the SD spectrum than in the absorption spectrum indicates the potential of the SD technique as a spectroscopy tool. We expect the fine structure in the SD spectrum to be specific to a particular chemical compound, which may lead to a nonlinear near-edge EUV spectroscopy technique based on SD. One can also envision looking for signatures of coherent FWM effects: for example, a photon echo can be detected in SD by introducing a delay between the two pulses [36, 37]. Even though we believe that the coherent $\chi^{(3)}$ response is unlikely to tangibly contribute to the SD signal in the present experiment, one can use absorption edges of lighter elements with longer dephasing time [5] and shorter EUV pulses. Experiments with gas phase samples, where hundreds-fs dephasing times have been reported [38] may prove especially promising. Whereas the present experiment was conducted at an FEL facility, the simplicity of the SD setup and large signal levels may enable "table-top" EUV FWM experiments with high-harmonic generation sources [39].

In summary, we have demonstrated femtosecond SD in the EUV range. By scanning the photons energy across the $M_{2,3}$ edge of Co, we obtain an SD spectrum revealing a resonant peak at the absorption edge, and a fine structure above the edge. The SD spectrum is much more structured than the EUV absorption spectrum, which may yield a useful nonlinear EUV spectroscopy tool. While a model based on an incoherent mechanism in which the SD signal is produced by a modulation of the electronic temperature yields a reasonable agreement with the experiment, it is anticipated that SD can be used to study coherent FWM effects in the EUV and possibly X-ray ranges.

**Acknowledgments**

A.A.M. and K.A.N. received support from the Department of Energy, Office of Science, Office of Basic Energy Sciences, under Award Number DE-SC0019126. W.L. and S.K.C. were supported as part of Ensembles of Photosynthetic Nanoreactors (EPN), an Energy Frontiers Research Center funded by the U.S. DOE, Office of Science under Award No. DE-SC0023431. W.L. acknowledges support from the Korea Foundation for Advanced Studies. D.K. and C.G. acknowledge funding by the Deutsche Forschungsgemeinschaft (DFG) projects GU 535/9-1 and KS 62/3-1. C.G. acknowledges funding from BMBF(05K24PSA) and DFG (NFDI 40/1). F.K. acknowledges funding by the Deutsche Forschungsgemeinschaft (DFG, German Research Foundation) TRR 173/2 Spin+X (Project A01 and B02). The calculations presented here were conducted in the Resnick High Performance Computing Center, a facility supported by Resnick Sustainability Institute at the California Institute of Technology.

# Resonant Self-Diffraction of Femtosecond Extreme Ultraviolet Pulses in Cobalt

## Supplemental Material

Alexei A. Maznev,[1,*] Wonseok Lee,[2] Scott K. Cushing,[2] Dario De Angelis,[3] Danny Fainozzi,[3] Laura Foglia,[3] Christian Gutt,[4] Nicolas Jaouen,[5,6] Fabian Kammerbauer,[7] Claudio Masciovecchio,[3] Riccardo Mincigrucci,[3] Keith A. Nelson,[1] Ettore Paltanin,[3] Jacopo Stefano Pelli-Cresi,[3] Vincent Polewczyk,[8] Dmitriy Ksenzov,[4] Filippo Bencivenga.[3]

[1]Department of Chemistry, Massachusetts Institute of Technology, Cambridge, Massachusetts 02139, USA
[2]Division of Chemistry and Chemical Engineering, California Institute of Technology, California 91125, USA
[3]Elettra-Sincrotrone Trieste, SS 14 km 163,5 in AREA Science Park, 34149 Trieste, Italy
[4]Department Physik, Universität Siegen, Walter-Flex-Strasse 3, 57072 Siegen, Germany
[5]Synchrotron SOLEIL, L'Orme des Merisiers, Saint-Aubin, Gif-sur-Yvette Cedex, 91192, France
[6]Department of Molecular Sciences and Nanosystems, Ca' Foscari University of Venice, 30172 Venezia, Italy
[7]Institute of Physics, Johannes Gutenberg University Mainz, 55099 Mainz, Germany
[8]Université Paris-Saclay, UVSQ, CNRS, GEMaC, 78000, Versailles, France
*Corresponding author: alexei.maznev@gmail.com

## CONTENTS

## S1. Electronic structure calculations for hcp cobalt

The electronic density of states (DOS) of cobalt (Co) is calculated within density functional theory (DFT) using Quantum ESPRESSO version 7.0 [1, 2]. Co adopts a hcp structure at ambient conditions with lattice constants of $a$ = 4.736 Bohr and $c$ = 7.693 Bohr [3]. A kinetic energy cutoff of 100 Ry is applied for the plane-wave expansion of the Kohn-Sham wavefunctions. The calculations use a norm-conserving Troullier-Martins pseudopotential [4] with the Perdew-Burke-Ernzerhof generalized gradient approximation exchange-correlation functional [5]. The integration of the Brillouin zone is performed using a 12 × 12 × 6 Monkhorst-Pack (MP) $k$-point grid [6], and Gaussian smearing is employed to accelerate self-consistent field convergence. The electronic DOS is computed with a 0.1 eV energy grid step. As shown in Fig. S1, the spin-up DOS is below the Fermi level $E_F$, with the $d$-band electrons being occupied, while the spin-down DOS cuts through $E_F$. These results are consistent with previous computational studies on hcp Co [7, 8].

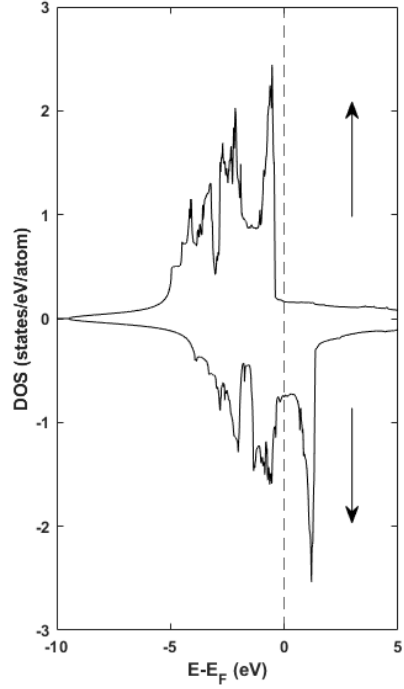

**FIG. S1.** Spin-polarized electronic DOS of hcp Co.

## S2. Chemical potential calculations

Given that the total number of valence electrons $N$ is conserved at any electronic temperature $T_e$, the chemical potential $\mu(T_e)$ is determined by integrating the product of the Fermi-Dirac distribution function, $f(E, \mu(T_e), T_e) = \{\exp[(E - \mu(T_e))/(k_B T_e)] + 1\}^{-1}$ with Boltzmann constant $k_B$, and the total electronic DOS of Co, $g(E)$, over all energies [9, 10]:

$$N = \int_{-\infty}^{\infty} f(E, E_F, T_e = 0 \text{ K}) g(E) dE = \int_{-\infty}^{\infty} f(E, \mu(T_e), T_e) g(E) dE. \qquad (1)$$

Co has $3d^7 4s^2$ valence electrons and exhibits a high electronic DOS near the Fermi level, as illustrated in Fig. S1. This high DOS allows the excitation of $d$-band electrons at the energy levels around the Fermi energy, which results in an increase in the chemical potential as $T_e$ increases, as shown in Fig. S2.

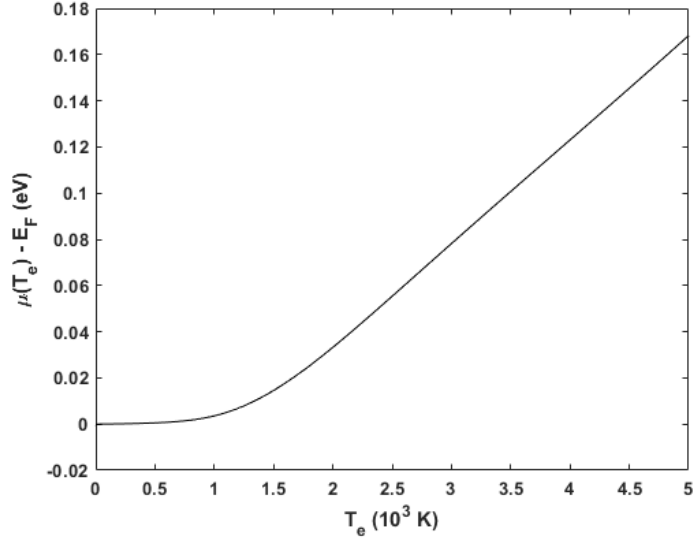

**FIG. S2.** Chemical potential of Co as a function of the electronic temperature.

### S3. Complex refractive index at varying $T_e$

Calculations of the complex refractive indices near the Co $M_{2,3}$ edge at varying $T_e$ are performed using the Obtaining Core Excitations from the *Ab initio* electronic structure and the NIST Bethe-Salpeter equation (BSE) solver (OCEAN) version 3.0.3 [11, 12]. The code first calculates the ground-state electronic structure of Co within plane-wave DFT framework using Quantum ESPRESSO and subsequently computes the complex dielectric function within the BSE approach to account for excitonic effects. For consistency, the same pseudopotential, kinetic energy cutoff, and lattice constants are used in both the DFT and BSE calculations. The MP $k$-point grids employed are 10 × 10 × 10 for the ground state, 16 × 16 × 16 for the final state, and 4 × 4 × 4 for screening calculations. The number of bands for the final-state and screening wavefunctions is set to 100, and a scaling factor of 0.8 is used for the Slater integrals, which is typical for 3$d$ transition metals [13]. The electron-hole occupation number differences in the BSE Hamiltonian are determined by the Fermi-Dirac distribution function, with chemical potentials dependent on $T_e$. The resulting dielectric functions are broadened using a convolution of a Lorentzian function with a FWHM of 0.3 eV. Finally, the complex dielectric function $\epsilon = \epsilon_1 + i\epsilon_2$ is converted into the complex refractive index $\tilde{n} = 1 - \delta + i\beta$ by using $1 - \delta = \{[(\epsilon_1^2 + \epsilon_2^2)^{1/2} + \epsilon_1]/2\}^{1/2}$ and $\beta = \{[(\epsilon_1^2 + \epsilon_2^2)^{1/2} - \epsilon_1]/2\}^{1/2}$. Figure S3 illustrates the experimental and calculated complex refractive indices of Co. While the calculations do not as accurately reproduce the experimental complex refractive index, the spectral differences due to photoexcitation is accurately reproduced as proven in previous studies [14, 15]. Figure S4 demonstrates that the differences in $\delta$ and $\beta$ resulting from a 100 K increase in $T_e$ are most pronounced near the Co $M_{2,3}$ edge, where they contribute to the most intense signal in the self-diffraction spectrum.

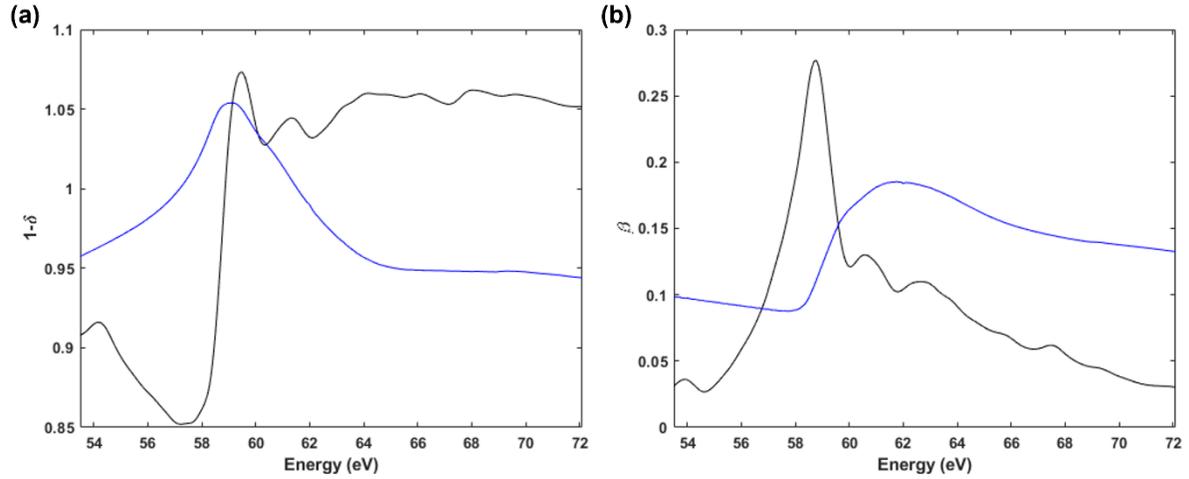

**FIG. S3.** (a) Real and (b) imaginary parts of the complex refractive index of Co, as obtained from [16] (blue) and from calculations (black).

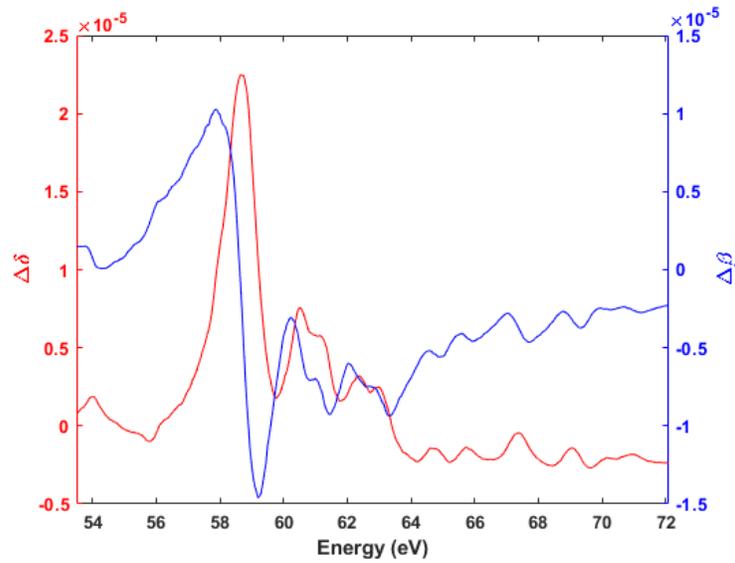

**FIG. S4.** $\Delta\delta$ (red) and $\Delta\beta$ (blue) at an electronic temperature rise of 100 K from 300 K to 400 K.